\documentclass[%
 reprint,
superscriptaddress,
 amsmath,amssymb,
 aps,
prb,
twocolumn
]{revtex4-2}

\usepackage{graphicx}% Include figure files
\usepackage{dcolumn}% Align table columns on decimal point
\usepackage{bm}% bold math
\begin{document}
\preprint{APS/123-QED}

\title{Universal work statistics in quenched gapless quantum systems} %Force line breaks with \\

\author{Donny \surname{Dwiputra}}
 \email{donny.dwiputra@apctp.org}
 \affiliation{Asia Pacific Center for Theoretical Physics, Pohang 37673, South Korea}
 \affiliation{Research Center for Quantum Physics, National Research and Innovation Agency (BRIN), South Tangerang 15314, Indonesia}
 
\author{Mir \surname{Faizal}}
\affiliation{Irving K. Barber Faculty of Science, University of British Columbia,
3333 University Way, Kelowna, British Columbia, Canada V1V 1V7}
\affiliation{Canadian Quantum Research Center 204-3002  32 Ave Vernon, BC V1T 2L7 Canada}

\author{Francesco \surname{Marino}}
\affiliation{Istituto Nazionale di Ottica del Consiglio Nazionale delle Ricerche (CNR-INO) c/o LENS, 50019 Sesto Fiorentino, Italy}
\affiliation{Istituto Nazionale di Fisica Nucleare, Sez. di Firenze, 50019 Sesto Fiorentino, Italy}

\author{Freddy P. \surname{Zen}}
\affiliation{Theoretical Physics Laboratory, Faculty of Mathematics and Natural Sciences, Institut Teknologi Bandung, Jl. Ganesha 10, Bandung 40132, Indonesia}
 \affiliation{Indonesian Center for Theoretical and Mathematical Physics (ICTMP), Bandung 40132, Indonesia}
\date{\today}

\begin{abstract}
We study the universality of work statistics performed during a quench in gapless quantum systems. We show that the cumulants of work scale separately in the fast and slow quench regimes, following a power law analogous to the universal scaling in the Kibble-Zurek mechanism for topological defect formation in phase transition. As an example, we analyze the nonequilibrium dynamics of a quenched Heisenberg XXZ chain at its critical gapless state using the bosonization picture, resulting in a Tomonaga-Luttinger liquid. The analytical scaling is in agreement with the exact numerical calculation for the fast and slow quench regimes. In finite systems, the characteristic function display an oscillatory pattern which disappears in the thermodynamic limit. This study is particularly useful for understanding the thermodynamics of adiabatic quantum computation.
\end{abstract}

\maketitle

\textit{Introduction.}---When a closed quantum system is driven across a second-order phase transition in a finite quench time $\tau_Q$, the equilibrium relaxation time diverges due to the gap closing. This critical slowdown implies that no matter how slowly a system is driven through the transition, its evolution cannot be adiabatic close to the critical point, leading to spontaneous nucleation of topological defects \cite{sengupta2004quench,dziarmaga2010dynamics,polkovnikov2011colloquium}. This nonequilibrium process is described by the Kibble-Zurek mechanism (KZM) \cite{kibble1976topology,kibble1980some,zurek1985cosmological,zurek1996cosmological,chandran2012kibble}, which predicts a universal scaling of the average defect density $n_\text{ex}\sim\tau_Q^{-\beta}$ \cite{del2014universality}. The power-law exponent $\beta=d\nu/(z\nu+1)$ is independent of the system details and fixed by the dimensionality of the system $d$ and by a combination of the correlation length and dynamic critical exponents, $\nu$ and $z$, respectively. Moreover, higher cumulants of $n_\text{ex}$ also exhibit the same scaling exponent for the 1D transverse Ising chain \cite{del2018universal}. However, this power-law scaling only holds for systems whose energy gap above the ground state follow a power-law opening near the critical points. Other universality classes, such as the Kosterlitz-Thouless transition, generally exhibit nonpower-law behavior that may only approach a power-law scaling asymptotically \cite{dziarmaga2014quench,gardas2017dynamics,zuo2021scaling}.

In small systems, where fluctuation is significant, nonequilibrium fluctuations are governed by the fluctuation theorems \cite{jarzynski2011equalities,campisi2011colloquium,vinjanampathy2016quantum}, which state that the work distribution for backward processes is exponentially suppressed. However, the notion of work in the context of fluctuation theorems for a Hamiltonian quantum system is subtle, since work is not observable \cite{talkner2007} or a single expectation value. Instead, quantum work involves two projective measurements within the two-time measurement (TTM) scheme, where one expects the full statistics of work done over all possible energy differences between the initial and final states. For quenches approaching or crossing the critical point in Ising-like systems, all cumulants obey power-law scaling with $\tau_Q$, albeit with cumulant-dependent exponent \cite{Fei:2020bjh,fei2019group,solfanelli2025universal}. Cumulants are generated by the characteristic function of work (CFW), which encodes essential information about nonequilibrium thermodynamics, such as the emergence of irreversibility (via fluctuation relations \cite{esposito2009nonequilibrium,seifert2012stochastic,dorner2012emergent}), and has a relation with the Loschmidt echo \cite{chenu2018quantum,chenu2019work,venuti2010unitary} as well as complexity \cite{pal2023time} in sudden quenches. The thermal average of the former quantity is equal to the out-of-time order correlator, a widely used measure of scrambling in chaotic quantum systems \cite{hashimoto2017out,swingle2018unscrambling}.

However, systems that have a gap above the ground state are exceptions rather than the rule. In reality, most systems with broken continuous symmetries have gapless excitations (Goldstone modes). Such as sound waves (phonons) in solid, magnons in ferromagnets and antiferromagnets, and Bogoliubov excitations in superfluids. Even in gapped systems with infinite size, we typically have a continuum spectrum above the gap, which is occupied by quasiparticles at finite temperatures. In gapless systems, the adiabatic limit breaks down, since any arbitrarily slow quenches excite the ground state. Despite the failure of adiabaticity, in low-dimensional gapless systems, there are regimes where the energy density is independent of the system size \cite{polkovnikov2008breakdown}, and thus adiabatic in slow quench limit. Moreover, adiabatic perturbation theory is still valid for extracting the scaling of quantities \cite{de2010quench}. In our example gapless system, the adiabatic quench is reached at $\tau_Q\to\infty$ in the sense that the CFW becomes a constant. However, the excitation probability $p_q$ shows a power-law decay instead of exponential as in the standard Landau-Zener (LZ) formula of an avoided crossing \cite{damski2005simplest}.

In this Letter, we study the work statistics of finite-rate quenched gapless systems in the spirit of the KZM mechanism. We show that the CFW in fast and slow quench regimes exhibit different scaling exponents and oscillating for finite systems. As an example system, we take the Heisenberg XXZ chain with quenched anisotropy parameter from its ground state. The system is mapped to a time-dependent Tomonaga-Luttinger liquid (TLL) using Abelian bosonization \cite{von1998bosonization,miranda2003introduction}. The analytical result is in good agreement with the exact numerical CFW of a finite XXZ chain, where both share common features for the fast and slow quench regimes.

%However, it is usually a highly challenging task to calculate the CFW for an interacting many-body systems with quenched parameters, even when the exact solution is at hand. Fortunately, an interacting (quartic) fermionic system can mapped into a quadratic bosonic Tomonaga-Luttinger liquid (TLL) using the bosonization technique \cite{miranda2003introduction,giamarchi2003quantum}. This method is versatile to describe the effective low-temperature behavior of a system in gapless state. Once the system is quadratic, one can use a general method to obtain the CFW under an arbitrary quench protocol by employing the group-representation theory in Ref. \cite{fei2019group}. The corresponding bosonic collective modes as the elementary excitation exhibit strong quantum fluctuations. How this system reacts to a time dependent protocol of finite-rate quench is a highly nontrivial problem, though some of its properties are already known \cite{dora2011crossover}. 

\textit{Work statistics following a quantum quench.}---The work distribution during the quench can be defined via the standard TTM scheme. For a generic isolated quantum many-body system prepared in an initial state $\rho$ the evolution from time $t=0$ to time $t=\tau_Q$ is described by a time-dependent Hamiltonian $H(t)$. The final state is thus $\rho(\tau_Q)=U\rho U^\dag$, where $U(\tau_Q)=\mathcal T \exp[-i\int_0^{\tau_Q} ds H(s)]$ (we set $\hbar=1$) and $\mathcal T$ is the time order operator. As long as the evolution is unitary, the changes in total energy from $H_0=H(0)$ to $H_\tau\equiv H(\tau_Q)$ can be properly identified as work \cite{alicki1979, frenzel2014, deffner2016, dahlsten2017, alipour2022}, even for mixed states. For an initial incoherent state (commutes with $H_0$), the instantaneous difference in energy $W_{nm}=\epsilon_n^\tau - \epsilon_m^0$ resulting from each pair of measurements is a stochastic quantity characterized by a probability distribution.
\begin{equation}
    P(W) = \sum_{m,n} \langle \epsilon_m^0|\rho|\epsilon_m^0\rangle p_{n|m} \delta(W-W_{nm}),
\end{equation}
where $p_{n|m}=|\langle \epsilon_n^\tau|U(\tau_Q)|\epsilon_m^0\rangle|^2$ and $|\epsilon_n^t\rangle$ is the instantaneous eigenstate of $H(t)$. From now on, the index $\tau$ indicates time-dependent quantities at time $\tau_Q$. For sudden quenches, $p_{n|m}=|\langle \epsilon_n^\tau|\epsilon_m^0\rangle|^2$, while in the adiabatic limit, $p_{n|m}=\delta_{n,m}$. Note that for our gapless case, the term adiabatic does not necessarily imply that there is no excitation in case of ground state quenches. Instead, it simply refers to the limit $\tau_Q\to\infty$. 

It is easier to calculate the cumulant generating function (CFW) of $P(W)$ \cite{talkner2007},
\begin{equation}
    G(u) = \int_0^{\tau_Q} dW \, P(W) e^{iuW} = \text{Tr}(e^{iuH_\tau^H}e^{-iuH_0}\rho). \label{CFW}
\end{equation}
where $H_t^H \equiv H^H(t) = U^\dag(t)H(t)U(t)$ is the Heisenberg picture Hamiltonian. The conjugate auxiliary variable $u$ is sometimes referred to as the second time of evolution in the TTM scheme \cite{silva2008statistics,chenu2018quantum}. The CFW shares the same amount of information with $P(W)$, and is connected via the Legendre-Fenchel transformation. The cumulants of work $\kappa_{n\geq 1}$ are defined as expansions of logarithm of the CFW,
\begin{equation}
    \ln G(u) = \sum_{n=1}^\infty \frac{(iu)^n}{n!} \kappa_n, \quad \kappa_n = i^{-n} \frac{d^n \ln G(u)}{du^n}\Bigg|_{u=0}.
\end{equation}
It is well known that the cumulants of a probability distribution are related to its moments. For example, the first cumulant is the average, $\kappa_1=\langle W\rangle$, and the second cumulant is the variance, $\kappa_2 = \langle W^2\rangle - \langle W\rangle^2$. Higher nonzero cumulants indicate a non-Gaussian distribution. 

%We consider a quench starting from the ground state $\rho(0)=|E_0\rangle\langle E_0|$ {\bf This sentence should be moved where necessary. Related to this, it is unclear whether the results refer to quenches initiated in the ground state or in a generic thermal state..}.

\textit{Scaling of fast quench.}---We first briefly discuss the behavior of the CFW where the quench rate is faster compared to all physical scales, i.e. $\tau_Q\ll \alpha^{-1}$. In the following example of the TLL system, the fastest rate $\alpha$ is the UV cutoff. We consider a linear ramp quench protocol of a system parameter $\lambda(t)\sim t/\tau_Q$ for $0\leq t\leq \tau_Q$ and the total Hamiltonian $H[\lambda(t)] = H_0 +\lambda(t)H_1$. The system is initiated in the ground state $|0\rangle \equiv |\epsilon_0^0\rangle$ of $H_0$.

When $\tau_Q\to 0$, the time evolution operator $U(t) = \mathcal{T} e^{-i\int_0^t ds H[\lambda(s)]}$ can be written as the following Dyson series,
\begin{eqnarray}
      U(t) = 1 - i\tau_Q \int_0^{t/\tau_Q} dw H[\lambda(w\tau_Q)] -\tau_Q^2\int_0^{t/\tau_Q}dw \nonumber \\
    \times\int_0^{w/\tau_Q} dw' H[\lambda(w\tau_Q)] H[\lambda(w'\tau_Q)] + O(\tau_Q^3).
\end{eqnarray}
The above series is written in a way that the integrands do not depend on $\tau_Q$. Thus, a Hermitian operator $A^H(\tau_Q) = U^\dagger(\tau_Q)A_0 U(\tau_Q)$ satisfies \cite{fei2021universal}
\begin{equation}
    \langle A \rangle_{\tau_Q} -\langle A \rangle_0 \sim \tau_Q^2,
\end{equation}
where $\langle A \rangle_0=\langle\epsilon_0^0|A|\epsilon_0^0\rangle$. Hence, for fast quenches, all cumulants satisfy the quadratic scaling, $\kappa_n - (\kappa_n)_0\sim\tau_Q^2$ ($n\geq 1$), where $(\kappa_n)_0$ is the cumulant for sudden quenching, $\tau_Q = 0$; the scaling is independent of the model. This universal behavior is analogous to the breakdown of the Kibble-Zurek scaling for fast quenches \cite{zeng2023universal}, where the defect statistics become insensitive to the quench rates and lose the power-law scaling due to a finite quench depth $\Delta_\text{f}$.

\textit{Slow quench.}---In the slow quench scenario, we can examine how all cumulants scale with the small but finite quench rate $\tau_Q^{-1}$ using the adiabatic perturbation theory \cite{polkovnikov2008breakdown,de2010quench}. Using the fact that the system is initially prepared at the ground state and assuming no additional Berry phase, the dimensionless excitation probability (whose integral over momenta is the average defect density) can be approximated as \cite{polkovnikov2005universal,rigolin2008beyond}
\begin{equation}
    p_{{q}} \approx \left|\int_\infty^\infty d\lambda\,\langle\vec q|\frac{d}{d\lambda}|0\rangle \, e^{i\tau_Q\int^\lambda d\lambda' \,\delta\epsilon_{q}(\lambda') } \right|^2.
\end{equation}
This quantity is central in calculating the CFW. We assume a uniform $d$-dimensional system with a relevant dispersion relation $\epsilon_{ q}(\lambda) \sim \lambda q^z$ where the momentum is $q=|\vec q|$ and $z$ is the dynamical exponent, as well as a general scaling of the instantaneous excitation $\delta\epsilon_{q}(\lambda) = \epsilon_{q}(\lambda) - \epsilon_{0}(\lambda) = q^a F(\frac{\epsilon_q}{q^a})$ and $\langle\vec q|\partial_\lambda|0\rangle=\lambda^{-1}G(\frac{\epsilon_q}{q^a})$ where $q^a$ indicates the higher-order gapless excitation term and $F(\chi), G(\chi)$ are model-dependent functions with universal tail $F(\chi)\sim\chi$, $G(\chi)\sim 1/\chi$  for $|\chi|\gg 1$ \cite{dziarmaga2010dynamics}. Note that $z=a$ in absence of the higher-order excitation term. In contrast to excitations near a phase transition, here the universal correlation length exponent $\nu$ does not appear since our quench protocol always keeps the system on a critical line. To obtain the scaling law, we need to remove the dimensionful quantity $\tau_Q$ from the exponential function by defining a new variable $\chi=\lambda q^{z-a}$ so that the exponential factor only depends on $q$ via the dimensionless combination $\tau_Q q^{2a-z}$ (times a dimensionless integral). This means that there exists a characteristic wave vector $\hat q\sim \tau_Q^{-1/(2a-z)}$ that is associated with the freeze-out length $\hat \xi \sim \hat q^{-1}$ in the KZM. This motivates us to introduce the following rescaled quantities,
\begin{equation}
    x = \lambda \tau_Q^{(a-z)/(2a-z)}, \quad y=q \tau_Q^{1/(2a-z)},
\end{equation}
such that $F(\lambda q^{z-a})=F(xy^{z-a})$ does not depend on $\tau_Q$.

For quenches (with arbitrary $\tau_Q$) initiated at the ground state of an infinite $d$-dimensional fermionic (bosonic) system $N\to\infty$, the logarithm of the CFW reads \cite{suppl,Fei:2020bjh,solfanelli2025universal}
\begin{eqnarray}\label{CFW_ground}
    \ln \frac{G(u)}{G_a(u)}=\pm N\int_0^\infty \frac{d^dq}{(2\pi)^d}\,\ln[1+p_q(e^{2iu\epsilon_q^{\tau}}-1)] \nonumber \\
    = \pm N \sum_{n=1}^\infty \frac{(-1)^{n+1}}{n}\int_0^\infty \frac{d^dq}{(2\pi)^d}\, p_q^n(e^{2iu\epsilon_q^\tau}-1)^n,
\end{eqnarray}
where $\ln G_a(u)\equiv Niu\mu = 2iuE_{\text g}= \mp\frac{Niu}{\pi}\int_0^\infty dq(\epsilon_q^{\tau} - \epsilon_q^0)$ is the cumulant CFW in the adiabatic limit. It is notable that the deformed ground state energy enters the first-order cumulant in this limit. By expanding $\ln G(u)$ into powers of $iu$, the cumulants of work are $\kappa_1 = N(\mu \pm \int \frac{d^dq}{(2\pi)^d}\, \epsilon_q^\tau p_q)$ and $\kappa_{n\geq 2} \approx \pm N\int \frac{d^dq}{(2\pi)^d}\, (\epsilon_q^{\tau})^n p_q$. Note that we have ignored the subleading terms $p_q^n$ for $n\geq 2$, which do not influence the scaling behavior of $\kappa_n$. Hence, the cumulants scale with $\tau_Q$ as follows,
\begin{equation}\label{scaling}
    |\kappa_n -\kappa_\mu| \approx N\int \frac{d^dq}{(2\pi)^d} (\epsilon_q^\tau)^n p_q 
    \propto \tau_Q^{-\theta_n}\int\frac{d^dy}{(2\pi)^d}|y|^{nz} p(y)
\end{equation}
where $\kappa_\mu = N\mu\delta_{n1}$, the exponent $\theta_n = \frac{d+nz}{2a-z}$, and
\begin{equation}\label{py}
    p(y) = \left| \int_{x_0}^{x_\tau} \frac{dx}{x}\, G(xy^{z-a}) \,e^{i\int^x dx' y^a F(x'y^{z-a})}\right|^2.
\end{equation}
Note that $n=0$ is identical to the defect density in KZM. For $\theta_n>2$, the integral in Eq.~(\ref{py}) diverges and the leading term comes from the high-energy contribution, which can be treated by introducing a UV regularization, resulting in $\kappa_n-\kappa_a\sim \tau_Q^{-2}$. For $\theta_n=2$ the integral is marginal and a logarithmic correction appears. Hence, we obtain
\begin{equation}\label{full_scaling}
    \kappa_n -\kappa_\mu \sim
    \begin{cases}
        \tau_Q^{-\theta_n}, & \theta_n<2, \\
        \tau_Q^{-2}\ln \tau_Q, & \theta_n=2, \\
        \tau_Q^{-2}, & \theta_n > 2.
    \end{cases}
\end{equation}
This will become more transparent in the calculation of the following example system.
 
We also want to address that the above scaling can also be derived from the LZ Hamiltonian for linear quenches. Suppose a gapless system has, or can be unitarily transformed to, the following effective two-band model \cite{divakaran2008quenching},
\begin{equation}
    h_q (t) = \frac{1}{2}
    \begin{pmatrix}
        r_q t/\tau_Q & s_q \\
        s_q & -r_q t/\tau_Q 
    \end{pmatrix},
\end{equation}
where $r_q\sim q^z$ describe the relevant dispersion relation and $s_q \sim q^a$ is the minimum excitation gap. The excitation probability for $\left|\frac{r_q t^2}{2\tau_Q}\right| \gg 1$ is given by the LZ formula (two-sided quench) as $p_q\approx e^{-\pi s_q^2 \tau_Q/(2 r_q)}$ \cite{damski2006adiabatic}. Thus, the scaling can be obtained via integration by substitution as $\kappa_n-\kappa_a \sim 
\int d^d q \, \epsilon_q^n p_q \sim \tau_Q^{-\theta_n}$, which is consistent with Eq.~(\ref{scaling}).

\textit{Example in the gapless phase of XXZ chain.}---We derive the CFW of a quenched 1D spin-$1/2$ XXZ chain as a prototype model to study interacting system in its gapless phase \cite{collura2015quantum,pollmann2013linear}. The system is described by Hamiltonian
\begin{equation}\label{XXZ}
    H(t) = J\sum_{j=1}^N \left[S_j^x S_{j+1}^x + S_j^y S_{j+1}^y + \Delta(t) S_j^z S_{j+1}^z\right],
\end{equation}
where $J>0$ is the constant coupling parameter denoting the energy scale, $S^\alpha_j$ ($\alpha=x,y,z$) is the spin-1/2 operator on the $j$-th site, and $\Delta(t)=\Delta_\text{f}\,t/\tau_Q$ for $0\leq t\leq \tau_Q$ is the linear quench protocol for interaction anisotropy. We have implicitly assumed periodic boundary conditions $S^\alpha_{N+1}=S^\alpha_N$. The model presents three different phases: a ferromagnetic phase for $\Delta<-1$, an antiferromagnetic (Néel) phase for $\Delta>1$, and a gapless TLL phase in the intermediate regime $-1 < \Delta \leq 1$ characterized by the power-law decay of correlations with $\Delta$-dependent exponent at large distances \cite{collura2015quantum}.

The effective gapless dynamics can conveniently be treated, after a Jordan-Wigner transformation to interacting fermions, using standard Abelian bosonization in the continuum limit \cite{miranda2003introduction}, yielding a quadratic TLL Hamiltonian,
\begin{equation}
H(t) = v_F \sum_{\eta,q>0} \omega_q(t) b_{q\eta}^{\dag}b_{q\eta} + \lambda_q(t)\left(b_{qR}^{\dag}b_{qL}^{\dag}+\text{H.c}\right), \label{H}
\end{equation}
where $v_F=J$ is the Fermi velocity, $b_\eta$ and $b^\dag_\eta$ ($\eta=L,R$) are the bosonic ladder operators, which create left- and right-moving density waves with wave number $q=2\pi n/N$, $n=1,\dots,N$. The time-dependent coefficients 
$\omega_q(t) = q (1+g_{4}(t))$ and $\lambda_q(t)=q g_2(t)$ are related to perturbative parameters $g_2$ and $g_4$ and, in turn, to the TLL parameters $v$ and $K$ by \cite{giamarchi2003quantum}
\begin{equation}
    vK=v_F(1+g_4-g_2), \quad \frac{v}{K}=v_F(1+g_4+g_2). \label{TL}
\end{equation}
For XXZ chains with $|\Delta(t)|<1$, the TLL parameters read \cite{sirker2006open,schoenauer2019finite}
\begin{equation}
    v=J  \frac{\pi\sqrt{1-\Delta}}{2\cos^{-1}\Delta}, \quad K=\frac{\pi/2}{\pi-\cos^{-1}\Delta}. \label{TLL}
\end{equation}

To calculate the excitation probability $p_q$, we consider the evolution of $b_q^H$ and map it to the diagonalized mode $d_q^H$ to obtain $p_q=|y_{2q}^\tau|^2$. The evolution of both are
\begin{eqnarray}
    b^H_{q R,L}(t) &=& x_{1q}(t)\,b_{q R,L} - x_{2q}^*(t)\,b_{q L,R}^\dag, \label{b}\\
    d_{q1,2}^H(t) &=& y_{1q}(t)\, d_{q1,2} + y_{2q}^*(t)\, d_{q2,1}^\dag.
\end{eqnarray}
Using $f_\pm(t) = x_1(t) \pm x_2(t)$ and the TLL parameters in Eq.~(\ref{TL}), we can decouple Eq.~(\ref{b}) and obtain a Sturm-Liouville type 
equation,
\begin{equation}
    \ddot f_-(t) - \frac{d}{dt}\ln\left[v(t)K(t)\right] \dot f_-(t) + (qv(t))^2 f_-(t) = 0,  \label{SL}
\end{equation}
and $f_+(t)=i\dot f_-(t)[qv(t)K(t)]^{-1}$. Eq.~(\ref{SL}) applies to general TLL and can be viewed as a generalization of the time evolution derived in Refs. \cite{bernier2014correlation,chudzinski2016time}. For a arbitrary $\tau_Q$ and small $|\Delta_\text{f}|\ll 1$, $vK$ becomes constant, and thus the evolution of the bosonic operator is Galilean invariant. In this limit, Eq.~(\ref{SL}) becomes $\ddot f_- +(Jq)^2(1+t/\tilde{\tau})f_-=0$. We perform a direct attack on this equation and obtain the exact solution in terms of Airy functions \cite{abramowitz1988handbook}. The asymptotics simplifies the slow quench excitation probability to \cite{suppl}
\begin{equation}
    p_q\approx p_0\,\text{sinc}^2(Jq\tau_Q),
\end{equation} where $\text{sinc}(x)=\sin(x)/x$ and $p_0 =(\Delta_\text{f}/\pi)^2$; this result agrees with Refs. \cite{dora2011crossover,dora2012generalized,bacsi2013quantum}. Note that $p_q$ decays algebraically with $q$ instead of exponentially as in typical noninteracting fermionic LZ problems \cite{dziarmaga2005dynamics,Fei:2020bjh}. We also numerically confirmed that, surprisingly, this probability is valid for all range of $\tau_Q$ as long as $|\Delta_\text{f}|$ is small.

Although the scaling in Eq.~(\ref{full_scaling}) applies for initial ground state, we extend the CFW calculation $G(u)$ for a thermal initial state, $\rho(0)=e^{-\beta H(0)}/Z(0)$; we will compare the scalings in both cases. We obtain \cite{suppl}
\begin{eqnarray}
    G(u) &=& e^{iuE_\text{g}^\tau}\prod_{q>0} \frac{g_q(u)}{g_q(0)}, \\
g_q(u) &=& \{1-\cos(u\epsilon_q^\tau)\cos\left[(u-i\beta)\epsilon_q^0\right] \nonumber \\
&& - \, Q_q \sin(u\epsilon_q^\tau)\sin\left[(u-i\beta)\epsilon_q^0\right]\}^{-1}
\end{eqnarray}
where $\epsilon^t_q=v^t q$ is the instantaneous dispersion relation and $Q_q = 1-2p_q$. The global phase comes from the ground energy difference $E_\text{g}^\tau$ in the Heisenberg picture. For ground state quenches ($\beta\to\infty)$, the logarithm of CFW becomes Eq.~(\ref{CFW_ground}) with a minus sign. As mentioned in the previous section, the scaling of $\kappa_n$ for slow quenches depends on $\int dq\, (\epsilon_q^\tau)^n p_q^m\propto\int dq\, q^k e^{-cq},\,c\in\mathbb C$, which diverges when $k=n-2m\geq -1$. The cumulant integral for the XXZ chain becomes \begin{equation}\label{integrals}
    \int_0^\infty dq\, (\epsilon_q^\tau)^n p_q^m\sim
    \begin{cases}
        \text{max}(\tau_Q^{-2m},\tau_Q^{-(n+1)}), & k>-1, \\
        \tau_Q^{-2m}\ln \tau_Q, & k=-1, \\
        \tau_Q^{-(n+1)}, & k < -1.
    \end{cases}
\end{equation}
We show these in the Supplemental Material \cite{suppl} by regulating UV divergences where $\alpha$ is the cutoff. Recall that only the leading order $m=1\,(k=n-2)$ determines the scaling. Thus, the above scaling agrees with Eq.~(\ref{full_scaling}) for $d=z=a=1$. The detailed results for the first cumulants are 
\begin{align}\label{cum_XXZ}
    \kappa_1 &= N\left[\mu - \int_0^{\infty}\frac{dq}{\pi}\,\epsilon_q^\tau p_q\right]= N\Bigg[\mu-\frac{v^\tau}{4\pi (J\tau_Q)^2} \nonumber\\
    &\times\ln\left(1+\frac{4J^2\tau_Q^2}{\alpha^2}\right)\Bigg]\,\sim\,\tau_Q^{-2}\ln\tau_Q,\nonumber \\
    \kappa_2 &= -\frac{2N}{\pi}\int_0^{\infty} dq \, \left(\epsilon_q^\tau\right)^2 p_q\left(1-p_q\right) \nonumber \\
    &= -\frac{N(v^\tau)^2p_0}{\pi(J\tau_Q)^2}\left[\frac{1}{\alpha}-\frac{\pi p_0}{2J\tau_Q}\right]\,\sim\, \tau_Q^{-2}.
\end{align}
Furthermore, higher cumulants include leading terms that fall into the first case of Eq.~(\ref{integrals}) so that higher cumulants saturate as $\kappa_{n>2}\sim\tau_Q^{-2}$. The fact that $\kappa_{n>2}$ is nonzero indicates that the work distribution $P(W)$ is non-Gaussian. In addition, by plugging $n=0$ we recover the KZM scaling for the average defect density $\langle n_\text{ex}\rangle\sim\tau_Q^{-1}$ as in Refs. \cite{del2018universal,del2014universality}.

\begin{figure*}
    \includegraphics[width=0.73\textwidth]{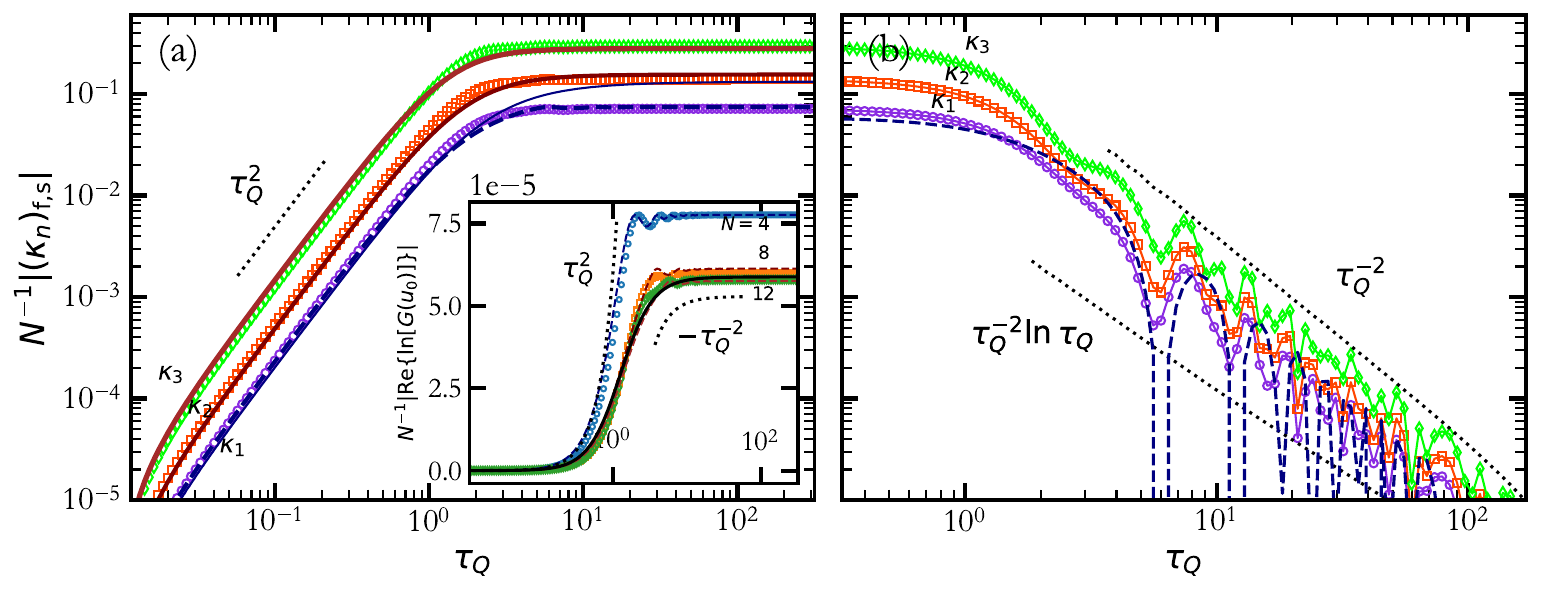}
    \caption{\label{fig:CFW} Renormalized cumulants $(\kappa_n)_\text{f,s}$ (absolute values) with respect to the (a) fast and (b) slow quench values as a function of the quench duration $\tau_Q$ (in units of $J^{-1}$). (a) Symbols represent exact numerics of a $N=12$ XXZ chain with $\Delta_\text{f}=0.1$. Solid lines are the analytical results in thermodynamic limit ($N=400$) fitted with UV cutoff $\alpha=3.51$. Dashed line indicates  finite size analytical $\kappa_1$. Inset: Real value of the full CFW for $N=(4,8,12)$ with $\alpha=(3.05,2.76,2.72)$ in linear-log plot. For slow quenches, the plateaus are approached according to power law $\tau_Q^{-2}$. The solid line represent the thermodynamic limit value. (b) The same data with (a) albeit renormalized with the adiabatic values (plateaus of the inset). Dashed line is the finite size analytical result of $\kappa_1$. Solid lines connecting the symbols are added to aid the eyes.}
\end{figure*}

To demonstrate the accuracy of these analytical results, we perform numerical simulations using the exact time evolution of a short XXZ chain quenched from the ground state of $\Delta=0$ to a small $\Delta_\text{f}$. Note that to correctly obtain the scaling behavior for fast and slow quenches, one needs to renormalize the cumulants with respect to their instantaneous, $(\kappa_n)_\text{f}$, and adiabatic, $(\kappa_n)_\text{s}$, values, respectively \cite{fei2021universal}. Figure \ref{fig:CFW}(a) shows the first three cumulants of a $N=12$ XXZ chain as a function of the quench duration $\tau_Q$. The three cumulants are shown to exhibit a universal $\tau_Q^{2}$ scaling, consistent with the fast quench prediction. For quenches slower than the system's dynamical time scale $\sim\Delta_\text{f}^{-1}$, all cumulants saturate to plateaus preceded with finite size oscillations, which are suppressed by system size (see inset). It is remarkable that the bosonization picture correctly predicted the finite size oscillations for small system sizes. On the other hand, the thermodynamic limit of $\kappa_1$ deviates from the values finite numerical and analytical values; it converges slower to the plateau. This due to the fact that the logarithmic correction in Eq.~(\ref{cum_XXZ}) only becomes significant in large systems. The KZM-like power-law scaling appears at the slow quench regime, once we renormalize with respect the plateau value (see the inset of Fig.~\ref{fig:CFW}(a)), as shown in Fig.~\ref{fig:CFW}(b) with $\tau_Q^{-2}$ scaling followed by the magnified oscillations due to the log-log plotting. Note that the $\tau_Q^{-2}$ and the $\tau_Q^{-2}\ln\tau_Q$ lines are the exact integration of Eq.~(\ref{CFW_ground}) for $N\to\infty$, rescaled to help the eyes. We also observe that as the system size $N$ grows, the scaling of $\kappa_1$ gradually shifts from $\tau_Q^{-2}$ to $\tau_Q^{-2}\ln\tau_Q$ with decreasing oscillations. The separation between fast and slow quenches scaling stems from the fact that the quench depth is finite and thus the power law scaling breaks down at the onset of adiabaticity.

\textit{Scaling of thermal quenches.}---For quenches initiated at a finite temperature $\beta^{-1}$, we obtain 
\begin{align}
    \kappa_1 &= N\int_0^{\infty} \frac{dq}{2\pi} \left( Q_q^\tau\epsilon_q^\tau- \epsilon_q^0 \right)\,\text{coth}\left(\frac{\beta\epsilon_q^0}{2}\right) \nonumber \,\sim \,(\beta\tau_Q)^{-1}, \\
    \kappa_2 &= \frac{N}{2} \int_0^{\infty} \frac{dq}{2\pi}\Big[\left(Q_q^\tau\epsilon_q^\tau-\epsilon_q^0\right)^2 - \left(1-(Q_q^{\tau})^2\right) (\epsilon_q^\tau)^2 \nonumber \\
    &\times\cosh(\beta\epsilon_q^0)\Big]\,\text{csch}^2\left(\frac{\beta\epsilon_q^0}{2}\right)\, \sim \, (\beta\tau_Q)^{-1}.
\end{align}
Evidently, all cumulants scales with $(\beta\tau_Q)^{-1}$ due to the bosonic bunching effect; the collective excitation of Luttinger liquid are bosons. In contrast, the scaling for thermal quenches of the transverse Ising-like models, whose quasiparticles are fermions, antibunching effect is expected \cite{Fei:2020bjh,suppl}.

\textit{Summary.}---In a gapless systems with a critical manifold, a divergence of the relaxation time result in the formation of topological defects as described by the Kibble-Zurek mechanism, whose main prediction is the universal scaling of the mean defect density (also in higher cumulants) with the quench duration. The imprint of KZM is also found thermodynamically in all the cumulants of quantum work statistics, albeit with different universal exponent due to the effect of dispersion relation. Since the quench is of a finite depth, there is a separation of scaling behavior between fast and slow regimes. We have confirmed these predictions in a quenched Heisenberg XXZ chain. In addition, we considered the scaling of cumulants for quenches initiated at a finite temperature. The scaling laws predicted here are directly testable in previously used platforms for KZM such as quantum annealers \cite{bando2020probing,king2022coherent}, or simulators of quantum systems with a gapless phase \cite{baier2016extended,yang2020observation}. This study can also be extended to a non-Hermitian systems \cite{dora2011crossover}.

\textit{Acknowledgements.}---DD is supported by the YST program at APCTP through the Science
and Technology Promotion Fund and Lottery Fund
of the Korean Government (and local governments
of Gyeongsangbuk-do Province and Pohang city). The many-body simulations are conducted using \texttt{QuTiP} package in \texttt{Python}~\cite{Johansson2012, Johansson2013}.

\bibliographystyle{apsrev4-2}
\bibliography{main_refs} 

\clearpage

\end{document}

\thispagestyle{empty}

\begin{center}
\large{\textbf{Supplemental material: \\ Universal work statistics in quenched gapless quantum systems}} 
\vspace{14pt}

\normalsize{Donny Dwiputra,$^{1, 2}$ Mir Faizal,$^{3, 4}$ Francesco Marino,$^{5, 6}$ and Freddy P. Zen$^{7, 8}$}
\vspace{12pt}

\begin{small}

$^1$\textit{Asia Pacific Center for Theoretical Physics, Pohang 37673, South Korea}

$^2$\textit{Research Center for Quantum Physics, National Research and Innovation Agency (BRIN), South Tangerang 15314,\nolinebreak Indonesia} 

$^3$\textit{Irving K. Barber Faculty of Science, University of British Columbia, \\ 3333 University Way, Kelowna, British Columbia, Canada V1V 1V7}

$^4$\textit{Canadian Quantum Research Center 204-3002  32 Ave Vernon, BC V1T 2L7 Canada}

$^5$\textit{Istituto Nazionale di Ottica del Consiglio Nazionale delle Ricerche (CNR-INO) c/o LENS, 50019 Sesto Fiorentino, Italy}

$^6$\textit{Istituto Nazionale di Fisica Nucleare, Sez. di Firenze, 50019 Sesto Fiorentino, Italy}

$^7$\textit{Theoretical Physics Laboratory, Faculty of Mathematics and Natural Sciences, \\ Institut Teknologi Bandung, Jl. Ganesha 10, Bandung 40132, Indonesia}

$^8$\textit{Indonesian Center for Theoretical and Mathematical Physics (ICTMP), Bandung 40132, Indonesia}
\end{small}
\end{center}

\vspace{3pt}

\section{Details of the time-dependent bosonization}
The quenched many-body system we consider in this Letter is the 1D spin-$1/2$ XXZ chain, which is described by the following Hamiltonian,
\begin{equation}\label{XXZ_S}
    H_\text{XXZ}(t) = J\sum_{j=1}^N \left[S_j^x S_{j+1}^x + S_j^y S_{j+1}^y + \Delta(t) S_j^z S_{j+1}^z\right],
\end{equation}
To bosonize the spin-$1/2$ XXZ chain, we first need to map it to the interacting spinless fermions using the Jordan-Wigner transformation. The Hamiltonian reads,
\begin{equation}
    H_\text{XXZ} = \frac{J}{2}\sum_j \Big(c_j^\dag c_{j+1}+\text{H.c.}\Big) + J\Delta\sum_j \Big(n_j-\frac{1}{2}\Big)\Big(n_{j+1}-\frac{1}{2}\Big) = H_{XY}+H_\text{int},
\end{equation}
where $n_j=c_j^\dag c_j$. The $H_{XY}$ part is the noninteracting part, corresponding to the $XY$ part of the spin-spin coupling. The periodic boundary conditions dictates that $\textbf{S}_{N+1}=\textbf{S}_1$, which result in a boundary term,
\begin{equation}
    S^x_N S^x_{N+1}+S^y_N S^x_{N+1} = \frac{e^{i\pi(K+1)}}{2}\Big(c^\dag_N c_1 + c_1^\dag c_N\Big),
\end{equation}
where $K$ is the total number of fermions. It is also convenient to transform the hopping term using a gauge transformation $c_j\to e^{i\pi j}c_j$ such that $c_j^\dag c_{j+1}\to-c_j^\dag c_{j+1}$ and $c_N^\dag c_1\to e^{i\pi(1-K)}c_N^\dag c_1$. The chain end hopping has now acquired a phase $e^{i\pi(K-L)}$. Note that the total magnetization is $M=\sum_j^N S_j^z = K-\frac{N}{2}$ \cite{miranda2003introduction}. Working on the zero-magnetization sector, $M=0$ and $K=\frac{N}{2}$, we can choose $N$ even, so that the end phase diseppears. In fermionic language this is $n_j=1/2$. The Fermi wave vector is $k_F=\pi \frac{K}{N} = \frac{\pi}{2}$, corresponding to a half-filled band, and the Fermi velocity becomes $v_F=J\sin k_F =J$.

We are now ready to bosonize the $XY$ part by linearizing around $\pm k_F$ to obtain two branches (right and left) of fermions running in $x\in(-\infty,\infty)$; we work in the thermodynamic (continuum limit). The fermionic field operator is
\begin{equation}\label{RL}
    \psi(x)\equiv e^{ik_F x}\psi_R(x)+e^{-ik_F x}\psi_L(x).
\end{equation}
The $XY$ part becomes the kinetic term ($:\;:$ denotes normal ordering),
\begin{equation}
    H_\text{XY}\equiv \sum_{q,\eta=L,R}q:c^\dag_{q\eta} c_{q\eta}:=\int dx\,\Big[:\psi_L^\dag(x)(i\partial_x)\psi_L(x):+:\psi_R^\dag(x)(-i\partial_x)\psi_R(x):\Big].
\end{equation}
This Hamiltonian can be bosonised by \cite{miranda2003introduction}
\begin{equation}
    \psi_{R,L} = \frac{F_{R,L}}{\sqrt{2\pi \alpha}}e^{-i\sqrt{2\pi}\phi_{R,L}}, \quad \phi_{R,L}(x) = \sum_{q>0}\frac{i e^{-\alpha q/2}}{\sqrt{Nq}}[e^{\pm iqx}b_{qR,L}-e^{\mp iqx}b^\dag_{qR,L}],
\end{equation}
where $F_{R,L}$ is the Klein factor and $\alpha$ is the UV cut-off. The bosonic fields satisfy $[\phi_{R,L}(x),\partial_y\phi_{R,L}(y)]=\mp i\delta(x-y)$ for $N\to\infty$. Up constant terms, the noninteracting Hamiltonian takes the following form,
\begin{equation}
    H_\text{XY} = \frac{v_F}{2}\sum_{\eta=R,L}\int_{-N/2}^{N/2} dx \, :(\partial_x\phi_\eta)^2: = \sum_{\eta=R,L} v_F q \,b^\dag_{q\eta}b_{q\eta}.
\end{equation}
Next, the interacting part we use $S_j^z=c_j^\dag c_j-\frac{1}{2}\approx:\psi^\dag(x)\psi(x):$ so that
\begin{equation}
    S_j^z S_{j+1}^z \approx \rho_R^2+\rho_L^2+4\rho_R\rho_L-\big(\psi^\dag_R\psi_L\big)^2 + \text{H.c.},
\end{equation}
where $\rho_\eta=:\psi_\eta^\dag\psi_\eta:=\mp\frac{1}{\sqrt{2\pi}}\partial_x\phi_{R,L}$. The last 2 terms are the umklapp (backscattering) terms and are irrelevant for $\Delta<1$. Thus the XXZ Hamiltonian reads,
\begin{eqnarray}
H_\text{XXZ}&\approx& \frac{v_F}{2}\int_{-N/2}^{N/2} dx \, \Big[(1+g_4)\sum_{q>0,\eta=R,L}:(\partial_x\phi_\eta)^2:\,-\,2g_2:(\partial_x\phi_R)(\partial_x\phi_L):\Big] \nonumber \\
&=& v_F\sum_{q>0,\eta=R,L}\Big[(1+g_4)b_{q\eta}^\dag b_{q\eta}+g_2(b_{qR}b_{qL}+\text{H.c.})\Big],
\end{eqnarray}
where $g_2=g_4=2\Delta/(2\pi v_F)$. Note that here the time dependence of the parameters is implicit. In the main paper we write the Hamiltonian as
\begin{equation}
\frac{1}{v_F} H_\text{XXZ}(t) =\sum_{\eta,q>0} \omega_q(t) b_{q\eta}^{\dag}b_{q\eta} + \lambda_q(t)\left(b_{qR}^{\dag}b_{qL}^{\dag}+\text{H.c}\right). \label{H_S}
\end{equation}

The parameters $g_2$ and $g_4$ are so far perturbative (only for $|\Delta|\ll 1$) since the effective bosonic Hamiltonian is derived by expanding the terms up to linear order, i.e., $c_{j+1}\to\psi(x+a)\approx \psi(x)+a\,\partial_x\psi(x)$ as implied by Eq.~(\ref{RL}) in $N\to\infty$, where $a$ is the lattice constant. Nevertheless, we can use the Hamiltonian for all $|\Delta|\leq 1$ by considering the Bethe Ansatz (BA) solution. BA allows to sum up exactly the infinite perturbative series for the excitation velocity \cite{sirker2006open}. To do this, we need to first diagonalize $H_{XXZ}$ using the bosonic Bogoliubov transformation,
\begin{equation}
    b_{q R(L)} = \cosh\gamma \,d_{q1(2)}-\sinh\gamma \,d^\dag_{q2(1)},
\end{equation}
where
\begin{equation}
    \gamma=\frac{1}{2}\ln K, \quad K=\Bigg(\frac{1+\Lambda}{1-\Lambda}\Bigg)^{1/2}=\sqrt\frac{1+g_4-g_2}{1+g_4+g_2},\quad \Lambda=\frac{g_2}{1+g_4}.
\end{equation}
The diagonalized Hamiltonian is, up to c-number terms (can be subtracted out by normal ordering),
\begin{equation}
    H_{XXZ} = v \sum_{q>0,i=1,2} q d_i^\dag d_i, \quad v = v_F\sqrt{(1+g_4)^2-g_2^2}.
\end{equation}
Thus we recover the following identities as in the main text,
\begin{equation}
    vK=v_F(1+g_4-g_2)=\omega-\lambda, \quad \frac{v}{K}=v_F(1+g_4+g_2)=\omega+\lambda.
\end{equation}
The correspondence with the exact solution obtained by BA allows us to redefine \cite{sirker2006open}
\begin{equation}\label{TLL_param}
    v = J\frac{\pi\sqrt{1-\Delta^2}}{2\cos^{-1} \Delta}=J\left(1+\frac{2\Delta}{\pi}+O(\Delta^2)\right), \quad K = \frac{\pi/2}{\pi-\cos^{-1}\Delta}=1-\frac{2\Delta}{\pi}+O(\Delta^2).
\end{equation}
The quench starts from $\Delta_0=0$ to a $\Delta_\text{f}$ at time $\tau_Q$, that is, $\Delta(t)=\Delta_\text{f}\,t/\tau_Q$. This induces a nontrivial time dependence of $v(t)$ and $K(t)$.

\section{Calculation of the CFW}
To calculate the CFW for a thermal initial state $\rho_0=e^{-\beta H_0}/Z_0$ with $Z_0=\text{Tr}(e^{-\beta H_0})$,
\begin{equation}\label{cfw}
    G(u) = \text{Tr}\left(e^{iuH^H(\tau_Q)}e^{-iuH_0}\rho_0\right),
\end{equation}
which contains the exponentials of $H(t)$, it is more convenient to work on the diagonalized basis (Bogoliubov) and to determine the Heisenberg evolution of the corresponding operators. Using the Bogoliubov transformations \cite{miranda2003introduction},
\begin{eqnarray}
     b_{q R,L} &=& \cosh \gamma(t)\, d_{q 1,2}(t) - \sinh \gamma(t) \, d_{q 2,1}^\dag(t), \\
     d_{q 1,2}(t) &=& \cosh \gamma(t)\, b_{q R,L} + \sinh \gamma(t)\, b_{q L,R}^\dag, \label{bogo}
\end{eqnarray}
where $\gamma(t) = -\frac{1}{2}\text{ln}\, K(t)$,
the Hamiltonian is diagonalized as follows,
\begin{equation}
    H(t) = E_\text{g}(t)+\frac{1}{2} \textbf{d}(t)^T \Lambda(t) \textbf{d}(t), \label{H_cast}
\end{equation}
where $E_g(t)=\sum_{q>0} [\epsilon_q(t)-\epsilon_q(0)]$ is the ground state energy, $\epsilon_q(t) = v(t)q$ are the instantaneous modal energies, $\textbf{d}(t) = \left(\{d_{q1}(t)\}, \{d_{q2}(t)\}, \{d_{q1}^\dag(t)\}, \{d_{q2}^\dag(t)\}\right)^T$ is $4N$-dimensional vector the components of which are Bogoliubov ladder operators relative to right-going and left-going density waves, with $\textbf{b}(t) = \left(\{b_{qR}(t)\}, \{b_{qL}(t)\}, \{b_{qR}^\dag(t)\}, \{b_{qL}^\dag(t)\}\right)^T$, and 
\begin{equation}
    \Lambda(t) =
    \begin{pmatrix}
        0 & 0 & [\epsilon(t)] & 0 \\
        0 & 0 & 0 & [\epsilon(t)] \\
        [\epsilon(t)] & 0 & 0 & 0 \\
        0 & [\epsilon(t)] & 0 & 0
    \end{pmatrix},
\end{equation}
where $[\epsilon(t)]$ is a diagonal $N\times N$ matrix of $\epsilon_q(t)$'s. 

Next, to find the Heisenberg-picture Hamiltonian $H^H(t)$, we need to solve the Heisenberg equation $\dot b^H_{q \eta}(t) =i[H(t),b^H_{q \eta}(t)]$. Using the fact that the quench creates excited modes for $t>0$,
\begin{equation}
    b^H_{q R,L}(t) \equiv x_{1q}(t)\,b_{q R,L} - x_{2q}^*(t)\,b_{q L,R}^\dag \label{b_coef},
\end{equation}
where $x_{1q}(0)=1$ and $x_{2q}(0)=0$, we obtain a couple of mode-independent differential equations for each $q$,
\begin{equation}
    i \dot x_{1,2}(t) = \pm\omega(t) \, x_{1,2}(t) \mp \lambda(t) \, x_{2,1}(t). \label{evo}
\end{equation}
Using $f_\pm(t) = x_1(t) \pm x_2(t)$ and the TLL parameters in Eq.~(\ref{TLL_param}), we can decouple Eq.~(\ref{evo}) to obtain a Sturm-Liouville equation,
\begin{equation}\label{ODE}
    \ddot f_-(t) - \frac{d}{dt}\ln\left[v(t)K(t)\right] \dot f_-(t) + (qv(t))^2 f_-(t) = 0,
\end{equation}
and $f_+(t)=i\dot f_-(t)[qv(t)K(t)]^{-1}$. The solution to the non-autonomous equation Eq.~(\ref{ODE}) is rather complicated and the exact solution for arbitrary $|\Delta_\text{f}|<1$ is currently unknown. We will introduce some approximations to treat this problem in the next section. One can also derive a similar equation of motion for an arbitrary initial $\Delta(0)$.

For now, we will continue to find $G(u)$, assuming that the solutions $f_{\pm}(t)$ are available at hand. The Heisenberg picture $H^H(t)$ can be readily written in its diagonal form,
\begin{equation}
    H^H(t) = E_\text{g}(t)+\frac{1}{2} \textbf{d}^H(t)^T \Lambda(t) \textbf{d}^H(t). \label{H_cast_heis}
\end{equation} 
We can resolve $d^H_{q1,2}(t)$ using the transformation in Eq.~(\ref{bogo}), and its inverse, in Heisenberg picture. Let us define
\begin{equation}
    d_{q1,2}^H(t) \equiv y_{1q}(t)\, d_{q1,2} + y_{2q}^*(t)\, d_{q2,1}^\dag,
\end{equation}
so that the solution can be written as $\textbf{d}^H(t) = \Upsilon(t) \textbf{d}(0)$, with
\begin{equation}
    \Upsilon(t) \equiv 
       \begin{pmatrix}
        [y_{1}(t)] & 0 & 0 &  [y_{2}^*(t)] \\
        0 &  [y_{1}(t)] &  [y_{2}^*(t)] & 0 \\
        0 &  [y_{2}(t)] &  [y_{1}^*(t)] & 0 \\
        [y_{2}(t)] & 0 & 0 &  [y_{1}^*(t)]
    \end{pmatrix},
\end{equation}
where $[y_{i}(t)]$ is a diagonal $N\times N$ matrix of $y_{iq}(t)$'s. Our aim is to relate $y_i(t)$'s to $x_i(t)$'s so that later we can obtain the excitation probability $p_q=|y_{q2}(\tau_Q)|^2$. Here, for each $q$, $y_{i}(t)$ is related to the coefficients $x_{1,2}(t)$ as
\begin{equation}
    \begin{pmatrix}
        y_{1}(t) & y_{2}^*(t)\\
        y_{2}(t) & y_{1}^*(t)
    \end{pmatrix}
    = 
    V(t) 
        \begin{pmatrix}
        x_{1}(t) & -x_{2}^*(t) \\
        -x_{2}(t)  & x_{1}^*(t)
    \end{pmatrix}
    V^{-1}(0), \label{transf}
\end{equation} 
with the canonical constraint $|y_{1}(t)|^2- |y_{2}(t)|^2 = 1$ and $|x_{1}(t)|^2- |x_{2}(t)|^2 = 1$, and the Bogoliubov transformation matrix
\begin{equation}
    V(t) = 
        \begin{pmatrix}
        \cosh \gamma(t) &\sinh \gamma(t) \\
        \sinh \gamma(t) &\cosh \gamma(t)
    \end{pmatrix}.
\end{equation}
The reason for this transformation between $y_i(t)$ and $x_i(t)$ will be clear in the next paragraph. For now we will list the corresponding transformation,
\begin{eqnarray}
    y_1(t) &=& \left[x_1(t)\cosh\gamma(t) - x_2\sinh\gamma(t)\right]\cosh\gamma(0) + \left[x_2^*(t)\cosh\gamma(t) - x_1^*(t)\sinh\gamma(t)\right]\sinh\gamma(0), \\
    y_2(t) &=& -\left[x_2(t)\cosh\gamma(t) - x_1\sinh\gamma(t)\right]\cosh\gamma(0) - \left[x_1^*(t)\cosh\gamma(t) - x_2^*(t)\sinh\gamma(t)\right]\sinh\gamma(0). \label{y2}
\end{eqnarray}
Here $x_{1,2}(t)=(f_+(t)\pm f_-(t))/2$ is obtained by solving Eq.~(\ref{ODE}).

From now on, we denote the superscript $\tau$ for quantities in time $t=\tau_Q$, e.g. $f_\pm^{\tau}=f_\pm(\tau_Q)$. Once the instantaneous energy $\epsilon_q(t)$ and $f^{\tau}_\pm$ is known, we need to relate the Heisenberg picture $\textbf{d}^H(t) = \Upsilon(t) \textbf{d}(0)$ to $\textbf b^H(t)=\textbf X(t) \textbf b(0)$ by relating $y_{1,2}(t)$ to $x_{1,2}(t)$. We can rewrite the Bogoliubov transformation as $\textbf{d}(t) = \textbf{V}(t)\textbf{b}(t)$, where
\begin{equation}
    \textbf{V}(t) =
    \begin{pmatrix}
        \textbf I\cosh \gamma(t) & \Sigma\sinh \gamma(t) \\
        \Sigma\sinh \gamma(t) & \textbf I\cosh \gamma(t)
    \end{pmatrix}, \quad
    \textbf{V}^{-1}(t) =
    \begin{pmatrix}
        \textbf I\cosh \gamma(t) & -\Sigma\sinh \gamma(t) \\
        -\Sigma\sinh \gamma(t) & \textbf I\cosh \gamma(t)
    \end{pmatrix}, \quad
    \Sigma=
    \begin{pmatrix}
        0 & I \\
        I & 0
    \end{pmatrix}_{2N\times 2N}.
\end{equation}
Also,
\begin{equation}
    \Upsilon(t) =
    \begin{pmatrix}
        y_1(t) \textbf I & y_2^*(t)\Sigma \\
        y_2(t)\Sigma & y_1^*(t)\textbf I
    \end{pmatrix}, \quad
    \textbf{X}(t)=
    \begin{pmatrix}
        x_1(t) \textbf{I} & -x_2^*(t)\Sigma \\
        -x_2(t)\Sigma & x_1^*(t)\textbf{I}
    \end{pmatrix}, \quad
    \Lambda(t)=
    \begin{pmatrix}
       0 & [\epsilon(t)] \textbf{I} \\
       [\epsilon(t)] \textbf{I} & 0
    \end{pmatrix}.
\end{equation}
Here, $I$ ($\textbf{I})$ is the $N\times N$ ($2N\times 2N$) identity matrix. Thus, $\textbf{d}^H(t)=\Upsilon(t)\textbf{d}(0)\iff\textbf{V}(t)\textbf{b}^H(t)=\Upsilon(t)\textbf{V}(0)\textbf{b}(0)\iff \textbf{V}(t)\textbf X(t)\textbf{b}(0)=\Upsilon(t)\textbf{V}(0)\textbf{b}(0)\iff\Upsilon(t) = \textbf{V}(t)\textbf{X}(t)\textbf{V}^{-1}(0)$.

The CFW with a thermal initial state is
\begin{equation}
    G(u) = \frac{\text{Tr}\big[e^{iuH_{\tau_Q}^H}e^{-i(u-i\beta)H_0}\big]}{\text{Tr}(e^{-\beta H_0})} = \frac{\text{Tr}\big[e^{\textbf{d}^T(iu\Upsilon^T_{\tau_Q} \Lambda_{\tau_Q} \Upsilon_{\tau_Q})\textbf{d}}\, e^{\textbf{d}^T [-i(u-i\beta)\Lambda_0]\textbf{d})}\big]}{\text{Tr}(e^{ \textbf{d}^T(-\beta\Lambda_0) \textbf{d}})}.
\end{equation}
To calculate the trace of the exponentials we use the trace formula for bosons in Eq.~(\ref{tr_formula}) \cite{fei2019group,munn1978ensemble},
\begin{equation}\label{tr_formula}
    \text{Tr}\left[\prod_{i} \exp\left(\frac{1}{2} \textbf{d}^T S_i \textbf{d}\right)\right] = \left[(-1)^n\left|\prod_{i} \exp(\tau_{\text B} S_i) - I\right|\right]^{-\frac{1}{2}},
    \quad \tau_\text{B}=
    \begin{pmatrix}
        0 & \textbf I \\
        -\textbf I & 0
    \end{pmatrix}.
\end{equation}
Thanks to $\tau_{\text B}$ and the structure of $\Lambda$, $\tau_\text{B}\Lambda$ is simply a diagonal matrix, while $\tau_B\Upsilon^T\Lambda\Upsilon$ is a $2\times 2$ block matrix. To calculate the exponential of these matrices, we can first perform diagonalization and use the decomposition $e^{A}=Se^\lambda S^{-1}$. Finally, we need to consider the following determinant formula for block matrices,
\begin{equation}
    \begin{vmatrix}
        A & B \\
        C & D 
    \end{vmatrix}=
    |A||D-CA^{-1}B|,
\end{equation}
provided that $A$ is invertible. If $[C,D]=0$ then the formula simply becomes $|AD-BC|$. In addition, the Sylvester's theorem holds, $|AB-I|=|BA-I|$.

In our case, there are $n=2N$ pairs of bosonic operators. Also, $S_1=iu\Upsilon^T_\tau\Lambda_{\tau_Q}\Upsilon_{\tau_Q}$ and $S_2=-i(u-i\beta)\Lambda(0)$ correspond to the exponentials containing $H_{\tau_Q}^H$ and $H_0$, respectively. The determinant can be calculated by considering the $n\times n$ block matrices following the matrix exponentials in Eq.~(\ref{tr_formula}). Hence, we obtain the CFW of the XXZ chain $G(u)$ for a thermal initial state,
\begin{equation}
    G(u) = e^{iuE_\text{g}^{\tau}}\prod_{q>0} \frac{g_q(u)}{g_q(0)},
\end{equation}
with
\begin{eqnarray}
g_q(u) &=& \{1-\cos(u\epsilon_q^{\tau})\cos\left[(u-i\beta)\epsilon_q^0\right] \nonumber \\
&& - \, Q_q^\tau \sin(u\epsilon_q^{\tau})\sin\left[(u-i\beta)\epsilon_q^0\right]\}^{-1}
\end{eqnarray}
where $\epsilon^{\tau}_q=v^{\tau} q$ is the instantaneous energy and $Q_q^{\tau} = 1-2|y_{2q}^{\tau}|^2$. Here $p_q = |y_{2q}^{\tau}|^2$ is the excitation probability, which is our central quantity in determining the scaling of the work statistics. This $p_q$ is obtained by solving $f^{\tau}_{\pm}$ in Eq.~(\ref{ODE}), finding $x_{1,2}^{\tau}$, and then mapping to $y_{1,2}^{\tau}$ via Eq.~(\ref{transf}).

For quenches initiated at the ground state ($\beta\to\infty)$, the cumulant CFW reads
\begin{eqnarray}
    \ln G(u) &=& N(iuE_\text{g}^\tau)-N\int_0^\infty \frac{dq}{2\pi}\ln\left[\frac{1}{2}\left(e^{iu\epsilon_q^\tau}+e^{-iu\epsilon_q^\tau}\right)e^{iu\epsilon_q^0} - \frac{Q_q^\tau}{2}\left(e^{iu\epsilon_q^\tau}-e^{-iu\epsilon_q^\tau}\right)e^{iu\epsilon_q^\tau}\right] \\
    &=& N(iuE_\text{g}^\tau) - N\int_0^\infty \frac{dq}{2\pi}iu(\epsilon_q^0-\epsilon_q^\tau) - N\int_0^\infty \frac{dq}{2\pi} \ln\left[1+p_q\left(e^{2iu\epsilon_q^\tau} - 1\right)\right].
\end{eqnarray}
In the main paper we write,
\begin{equation}
    \ln \frac{G(u)}{G_a(u)}=-N\int_0^\infty \frac{dq}{2\pi}\ln[1+p_q(e^{2iu\epsilon_q^{\tau}}-1)], \label{G_ground}
\end{equation}
where $\ln G_a(u)\equiv Niu\mu = 2iuE_\text{g}=Niu(\pi)^{-1}\int_0^\infty dq(\epsilon_q^{\tau} - \epsilon_q^0)$ is the cumulant CFW in the adiabatic limit. Note that we have used the continuum approximation $\frac{2\pi}{N}\sum_q\to\int_0^\infty dq$ so that a UV cutoff factor $e^{-\alpha q}$ must be included whenever the integral diverges. Note that to fit with the numerical data we need to adjust a proportionality factor to compensate the applied cutoff factor. On the other hand, to compare with the bosonic case, the logarithm of the CFW for quadratic fermions only differ in the overall sign. It reads as follows,
\begin{equation}
     \ln \frac{G(u)}{G_a(u)}=N\int_0^\infty \frac{dq}{2\pi}\ln[1+p_q(e^{2iu\epsilon_q^{\tau}}-1)], 
\end{equation}
where $\ln G_a(u)=Niu(\pi)^{-1}\int_0^\infty dq(\epsilon_q^{0} - \epsilon_q^\tau)$.

\section{The excitation probability for slow quenches}
Eq.~(\ref{ODE}) is substantially difficult to solve for an arbitrary interaction strength in the gapless regime, $|\Delta_\text{f}|\leq 1$. To proceed analytically, we need to make approximations $|\Delta_\text{f}|\ll 1$. For a sufficiently small $\Delta_\text{f}$, $vK$ becomes a constant and thus the evolution of the bosonic operator is Galilean invariant. In this limit, Eq.~(\ref{ODE}) becomes $\ddot f_- +(Jq)^2(1+t/\tilde{\tau})f_-=0$ and the solutions $f_\pm$ are in terms of Airy functions \cite{abramowitz1988handbook}. Note that we can also map the Airy functions into Bessel functions \cite{abramowitz1988handbook}. Here, we choose to work with the Airy functions because of the nice asymptotics. The solution at the end of the quench reads,
\begin{equation}
    f_-^{\tau} = \pi\big[(\text{Bi}'(\alpha) + \alpha^{1/2}\text{Bi}(\alpha))\text{Ai}(\alpha\kappa)-(\text{Ai}'(\alpha) + \alpha^{1/2}\text{Ai}(\alpha))\text{Bi}(\alpha\kappa)\big],
\end{equation}
where $\alpha = (iJq\tilde{\tau})^{2/3}$, $\tilde\tau=\tau_Q\pi/(4\Delta_\text{f})$, and $\kappa=1+4\Delta_\text{f}/\pi$. 

For slow quenches, with large enough $\tau_Q$ (large $|\alpha|$), one can perform the asymptotic expansion to get
\begin{equation}
    f_-^{\tau} = \kappa^{-\frac{1}{4}}\Bigg\{e^{-isq}\Bigg[1-\frac{1+5\kappa^{-3/2}}{48iJq\tilde\tau}\Bigg] + \frac{e^{isq}}{8iJq\tilde\tau}\Bigg\} + O\Big(\frac{1}{\tilde\tau^2}\Big),
\end{equation}
where $s=J\pi\tau_Q(6\Delta_\text{f})^{-1}\big[\kappa^{3/2}-1\big]$. Similarly, $f_+(t)=i\dot f_-(t)/(Jq)$,
\begin{eqnarray}
     f_+^{\tau} &=& -\frac{\pi}{\alpha^{1/2}}\big[(\text{Bi}'(\alpha) + \alpha^{1/2}\text{Bi}(\alpha))\text{Ai}'(\alpha\kappa)-(\text{Ai}'(\alpha) + \alpha^{1/2}\text{Ai}(\alpha))\text{Bi}'(\alpha\kappa)\big] \\
     &=&  \kappa^{\frac{1}{4}}\Bigg\{e^{-isq}\Bigg[1-\frac{1-7\kappa^{-3/2}}{48iJq\tilde\tau}\Bigg] - \frac{e^{isq}}{8iJq\tilde\tau}\Bigg\} + O\Big(\frac{1}{\tilde\tau^2}\Big).
\end{eqnarray}
We can also approximate $x_{1,2}^{\tau}=\frac{1}{2}(f_+^{\tau}\pm f_-^{\tau})$ and relate to the excitation probability $p_q=|y_2^{\tau}|^2$ where
\begin{equation}
    y_2^{\tau}=x_1^{\tau}\sinh \gamma_\tau - x_2^{\tau} \cosh \gamma_\tau.
\end{equation}
is obtained from Eq.~(\ref{y2}). Next, we expand $\kappa^a\approx1+4a\Delta_\text{f}/\pi$ terms to get
\begin{eqnarray}
    x_1^\tau &=& e^{-iJq\tau_Q} + O\Bigg(\frac{\Delta^2_\text{f}}{\tau_Q^2}\Bigg), \\
    x_2^\tau &=& \frac{\Delta_\text{f}}{\pi}\Bigg(e^{-iJq\tau_Q}-\frac{\sin(Jq\tau_Q)}{Jq\tau_Q}\Bigg) + O\Bigg(\frac{\Delta^2_\text{f}}{\tau_Q^2}\Bigg), \\
    y_2^\tau &\approx& \frac{\Delta_\text{f}}{\pi}\text{sinc}\left(Jq\tau_Q\right).
\end{eqnarray}
In the above, we have used $\sinh \gamma_\tau \approx \frac{\Delta_\text{f}}{\pi}$, $\cosh \gamma_\tau \approx 1$, and $s\approx J\tau_Q$ for $|\Delta_\text{f}|\ll 1$. This results in the excitation probability, 
\begin{equation}
    p_q = p_0 \;\text{sinc}^2\left(Jq\tau_Q\right),
\end{equation}
where $p_0 = (\Delta_\text{f}/\pi)^2$, as used in the main text. We have confirmed numerically that this approximate probability of excitation is valid for all $\tau_Q$ as long as $|\Delta_\text{f}|\ll 1$. 

\section{Scaling of cumulants for the ground state and thermal quenches}
In the following, we first calculate cumulants $\kappa_n$ for the quench initiated in the ground state of $H(0)$. We derive the important integrals, $\int_0^\infty dq\,(\epsilon_q^\tau)^n p_q^m$ with $n,m\in\mathbb Z^+$, which appear in the cumulant expansion. Recall that $\epsilon^{\tau}_q = v q$, where $v\equiv v^\tau$, and $p_q = p_0\,\text{sinc}^2(J\tau_Q)$. Note that the following integrals must be understood as the Cauchy principal value integral. Using the expansion,
\begin{equation}
    \sin^{2m}\theta = \frac{1}{2^{2m}}\binom{2m}{m}+\frac{(-1)^m}{2^{2m-1}}\sum_{k=0}^{m-1}(-1)^k\binom{2m}{k}\cos[2(m-k)\theta],
\end{equation}
we obtain the following integral, for $n-2m>-1$,
\begin{eqnarray}\label{int}
    &&\int_0^\infty dq\,(\epsilon_q^\tau)^n p_q^m \nonumber \\
    &&= \lim_{\alpha\to0}\int_0^\infty dq \,e^{-\alpha q}(\epsilon^\tau_q)^n p_q^m \nonumber \\
    &&=  \lim_{\alpha\to0}\frac{v^n p_0^m}{(J\tau_Q)^{1+n}}\int_0^\infty d\theta \, e^{-\alpha\theta/(J\tau_Q)}\theta^{n-2m}\sin^{2m}\theta \nonumber\\
    &&= \frac{v^np_0^m \Gamma(1+n-2m)}{2^{2m}}\left[\frac{\binom{2m}{m}}{\alpha^{1+n-2m}(J\tau_Q)^{2m}}-2\frac{(-1)^m}{(J\tau_Q)^{1+n}}\sum_{k=0}^{m-1}(-1)^k\binom{2m}{k}\frac{\sin\left(\frac{\pi}{2}(n-2m)\right)}{[2(m-k)]^{1+n-2m}}\right]. 
\end{eqnarray}
In the last line, we have used $i^{-1-l}+(-i)^{-1-l}=-2\sin(\pi l/2)$, with $l=n-2m$, which tells us that the second term vanishes for $l$ even. The first term is the cutoff-dependent term that scales with $\sim\tau_Q^{-2m}$, while the second term is independent of the cutoff and scales with $\sim\tau_Q^{-(n+1)}$. Note that for $n-2m=1$ a logarithmic correction will appear, as shown in $\kappa_1$ below, as well as in the $\int_0^\infty dq\, (\epsilon_q^\tau)^3p_q^2$ term of $\kappa_3$. For $n-2m=-1$,
\begin{eqnarray}\label{eq-1}
    &&\int_0^\infty dq \,(\epsilon^\tau_q)^n p_q^m \nonumber \\
    &&= \lim_{\alpha\to0} \frac{v^{-1+2m}p_0^m}{(J\tau_Q)^{2m}}\int_0^\infty d\theta\,e^{-\alpha\theta} \frac{\sin^{2m}\theta}{\theta} \nonumber \\
    &&= \lim_{a\to0}\lim_{\varepsilon\to0}\frac{v^{-1+2m}p_0^m}{(2J\tau_Q)^{2m}}\int_\varepsilon^\infty \left[ E_1(\alpha\varepsilon)\binom{2m}{m} + \sum_{k=0}^{m-1} (-1)^{m+k}\left(E_1\left([\alpha-2i(m-k)\varepsilon]\right) + \text{c.c.}\right)\binom{2m}{k} \right] \nonumber \\
    &&=-\frac{v^{-1+2m}p_0^m}{(2J\tau_Q)^{2m}}\left[\ln\left((\alpha\varepsilon)^{\binom{2m}{m}}\prod_{k=0}^{m-1}\left[(a^2+4(m-k)^2)\varepsilon^2\right]^{(-1)^{m+k}\binom{2m}{k}}\right) + \gamma\left(\binom{2m}{m}+2\sum_{k=0}^{m-1}(-1)^{m+k}\binom{2m}{k}\right)\right] \nonumber \\
    &&=\frac{v^{-1+2m}p_0^m}{(2J\tau_Q)^{2m}} \ln\left[(\alpha\varepsilon)^{\binom{2m}{m}\left(1-(-1)^m\right)}\prod_{k=0}^{m-1}\left(1+\frac{4(m-k)^2}{\alpha^2}\right)^{(-1)^{m+k}\binom{2m}{k}}\right] \nonumber \\
    &&= \frac{v^{-1+2m}p_0^m}{(2J\tau_Q)^{2m}} \sum_{k=0}^{m-1} \binom{2m}{k}(-1)^{m+k}\ln\left(1+\frac{4\left[(m-k)J\tau_Q\right]^2}{\alpha^2}\right)\; \sim\;\tau_Q^{-2m}\ln \tau_Q,
\end{eqnarray}
where $a=\alpha/(J\tau_Q)$ and we have approximated the exponential integral for small $z$, $E_1(z)=\int_z^\infty \frac{e^{-t}}{t}dt\approx -\ln z-\gamma$, where $\gamma$ is the Euler-Mascheroni constant. Note that in the fourth line $\sum_{k=0}^{m-1}(-1)^{m+k}\binom{2m}{k}=-\frac{1}{2}(-1)^{2m}$ so that $\varepsilon$ cancels in the fifth line. The last similarity shows the scaling behavior for $J\tau_Q\gg 1$. For $n-2m<-1$, the cut-off factor is not needed, since the integral already converges. That is, for $n-2m<-1$,
\begin{eqnarray}
    &&\int_0^\infty dq \,(\epsilon^\tau_q)^n p_q^m \nonumber \\
    &&= \lim_{\varepsilon\to 0}\frac{v^n p_0^m}{(J\tau_Q)^{1+n}}\int_\varepsilon^\infty d\theta\, \frac{\sin^{2m}\theta}{\theta^{p}} \nonumber \\
    &&= \lim_{\varepsilon\to 0} \frac{v^n p_0^m}{2^{2m}(J\tau_Q)^{1+n}} \left\{-\left[\frac{\theta^{-(p-1)}\binom{2m}{m}}{p-1}\right]_\varepsilon^\infty + 2\sum_{k=0}^{m-1}(-1)^{m+k}\binom{2m}{k}\int_\varepsilon^\infty d\theta\,\frac{\cos\left[2(m-k)\theta\right]}{\theta^p} \right\} \nonumber\\
    &&= \lim_{\varepsilon\to 0} \frac{-v^n p_0^m}{2^{2m}(J\tau_Q)^{1+n}}\left[\frac{\theta^{-(p-1)}\binom{2m}{m}}{p-1} + 2\sum_{k=0}^{m-1}(-1)^{m+k}\binom{2m}{k}\left(\sum_{j=0}^{p-2}(-1)^j\frac{q^j\cos\left(q-\frac{j\pi}{2}\right)}{\binom{p-1}{j}j!\,\theta^{p-1-j}}+\frac{q^{p-1}}{(p-1)!}
    \begin{cases}
        -\text{Si}(q\theta),\;p\,\text{even} \\
        \text{Ci}(q\theta), \;p\,\text{odd}
    \end{cases}\right)
    \right]_\varepsilon^\infty \nonumber \\
    &&= \frac{2v^n p_0^m}{2^{2m}(J\tau_Q)^{1+n}(2m-n-1)}\sum_{k=0}^{m-1}(-1)^{m+k}\binom{2m}{k}\left[2(m-k)\right]^{2m-n-1}
    \begin{cases}
        \frac{\pi}{2},\;p\;\text{even} \\
        \ln\left[2(m-k)\right], \;p\;\text{odd}
    \end{cases} \sim\;  \tau_Q^{-(n+1)},
\end{eqnarray}
where $p=2m-n$ and $q=2(m-k)$, and $\text{Si(Ci)(x)}$ is the sine (cosine) integral. To obtain the fourth line, we have used integration by parts. The coefficients of $1/\varepsilon^l$ cancel each other due to the binomial coefficients, similar to Eq.~(\ref{eq-1}). It is easier to show this via mathematical induction. In the last line, we have used $\text{Si}(\infty)=\pi/2$ and $\lim_{\varepsilon\to0}\text{Ci}(\varepsilon) = \ln\varepsilon+\gamma$. As an example, here we list some of the results,
\begin{equation}
    \int_0^\infty d\theta\;\frac{\sin^4 \theta}{\theta^2} = \frac{\pi}{4}, \quad  \int_0^\infty d\theta\;\frac{\sin^6 \theta}{\theta^3} = \frac{3}{16}\ln\left(\frac{256}{27}\right).
\end{equation}

Now, we have all the formula needed to calculate $\kappa_n$. Listed below are the cumulants for a quench initiated in the ground state. For $\kappa_1$, logarithmic correction appears as predicted by the scaling law in the main text. For the cumulants $n\geq 2$, since $\kappa_n$ always contains a $\int_0^\infty dq\,(\epsilon_q^\tau)^n p_q$ term, the leading order always scales like $\sim \tau_Q^{-2}$ for$J\tau_Q\gg1$.

\begin{eqnarray}
    \kappa_1 &=& N\left[\mu - \int_0^{\infty}\frac{dq}{\pi}\,\epsilon_q^\tau p_q\right]=N\left[\mu-\frac{v^\tau}{4\pi (J\tau_Q)^2}\ln\left(1+\frac{4J^2\tau_Q^2}{\alpha^2}\right)\right]\,\sim\,\tau_Q^{-2}\ln\tau_Q, \\
    \kappa_2 &=& -\frac{2N}{\pi}\int_0^{\infty} dq \, \left(\epsilon_q^\tau\right)^2 p_q\left(1-p_q\right) = -\frac{N(v^\tau)^2p_0}{\pi(J\tau_Q)^2}\left[\frac{1}{\alpha}-\frac{\pi p_0}{2J\tau_Q}\right]\,\sim\, \tau_Q^{-2}, \\
    \kappa_3 &=& -\frac{4N}{\pi}\int_0^{\infty} dq \, \left(\epsilon_q^\tau\right)^3 p_q\left(1-p_q\right)\left(1-2p_q\right)\,\sim\,\tau_Q^{-2}.
\end{eqnarray}
That is, they match the scaling law derived in the main text.

Now, we calculate the cumulants for the thermal initial state $\rho(0)=\frac{e^{-\beta H(0)}}{Z(0)}$. To proceed analytically, we consider the high temperature limit $\beta\to0$, that is, $\coth(x)\approx \frac{1}{x}$, $\text{csch}(x)\approx \frac{1}{x}$, and $\cosh(x)\approx 1$. Recall that $Q_q^\tau = 1-2p_q$. Using the results above, we get the scaling of the cumulants,
\begin{eqnarray}
    \kappa_1 &=& -N\int_0^{\infty} \frac{dq}{2\pi} \left(\epsilon_q^0 - Q_q^\tau\epsilon_q^\tau\right)\,\text{coth}\left(\frac{\beta\epsilon_q^0}{2}\right) \,\sim \,(\beta\tau_Q)^{-1}, \\
    \kappa_2 &=& - \frac{N}{2} \int_0^{\infty} \frac{dq}{2\pi}\left[\left(\epsilon_q^0-Q_q^\tau\epsilon_q^\tau\right)^2 + \left(1-(Q_q^{\tau})^2\right) (\epsilon_q^\tau)^2\cosh(\beta\epsilon_q^0)\right]\,\text{csch}^2\left(\frac{\beta\epsilon_q^0}{2}\right)\, \sim \, (\beta\tau_Q)^{-1}.
\end{eqnarray}
We can also show that $\kappa_3 \sim (\beta\tau_Q)^{-1}$. Higher cumulants $\kappa_{k>2}$ appear to have the same scaling, as thermal factors (hyperbolic functions) become $\sim 1/(\beta q)^k$ and the integral falls into the $n-2m<-1$ class with $n=0$ so that the scaling saturates to $\sim 1/(\beta\tau_Q)$. This scaling law is related to the bosonic bunching effect, since the collective excitations in the Luttinger liquid are bosons, in contrast to the fermionic antibunching effect \cite{de2010quench}. As for the 1D transverse Ising model, whose quasiparticles are fermions, the scaling is $\sim \beta /\tau_Q$ for odd cumulants and $\sim N \tau_Q^{-1}$ for even cumulants; see the supplemental material of Ref. \cite{Fei:2020bjh}. At zero temperatures and far from criticality, all slow-quenching cumulants in the Ising model scale with $\tau_Q^{-1/2}$ \cite{Fei:2020bjh}, which is half its high-temperature scaling. This is related to the fermionic antibunching effect at finite temperature.

\bibliographystyle{apsrev4-2}
\bibliography{supp_refs}